\documentclass[sigconf]{acmart}
\usepackage{booktabs}
\usepackage{multirow}
\graphicspath{{Figures/}}
\AtBeginDocument{%
  }

\begin{document}

\title{Execution-Centric Characterization of FP8 Matrix Cores, Asynchronous Execution, and Structured Sparsity on AMD MI300A}
% \title{Microbenchmark-Driven Analysis of Unified Memory, Coherence, and Interference on AMD MI300A}

\author{Aaron Jarmusch}
\orcid{1234-5678-9012}
\affiliation{%
  \institution{University of Delaware}
  \city{Newark}
  \state{Delaware}
  \country{USA}
}

\author{Connor Vitz}
\affiliation{%
  \institution{University of Delaware}
  \city{Newark}
  \state{Delaware}
  \country{USA}
}

\author{Sunita Chandrasekaran}
\affiliation{%
  \institution{University of Delaware}
  \city{Newark}
  \state{Delaware}
  \country{USA}
}

\renewcommand{\shortauthors}{Jarmusch et al.}

\begin{abstract}

The AMD MI300A APU integrates CDNA3 GPUs with high-bandwidth memory and advanced accelerator features: FP8 matrix cores, asynchronous compute engines (ACE), and 2:4 structured sparsity. These capabilities are increasingly relied upon by modern HPC and HPC-AI workloads, yet their execution characteristics and system-level implications remain insufficiently understood. In this paper, we present an execution-centric characterization of FP8 matrix execution, ACE concurrency, and structured sparsity on MI300A using targeted microbenchmarks. We quantify occupancy thresholds, fairness, throughput trade-offs under concurrent execution, and context-dependent sparsity benefits. We evaluate representative case studies—transformer-style, concurrent, and mixed-precision kernels—to show how these effects translate into application-level performance and predictability. Our results provide practical guidance for occupancy-aware scheduling, concurrency decisions, and sparsity enablement on MI300A-class unified nodes.

\end{abstract}



\received{4 February 2026}


\maketitle

\section{Introduction}
\label{sec:introduction}
Emerging heterogeneous systems increasingly rely on tightly integrated accelerators to deliver performance and energy efficiency for scientific computing and machine learning workloads. The AMD MI300A APU exemplifies this trend by combining high-core-count CPUs and CDNA3 GPUs within a single package, coupled with shared high-bandwidth memory and advanced accelerator features. Beyond unified memory, MI300A introduces support for FP8 matrix computation, asynchronous compute engines (ACE), and structured sparsity—features that are critical to modern HPC-AI workloads but whose practical performance characteristics remain opaque.

Prior work has focused primarily on characterizing unified memory behavior and CPU–GPU interference on MI300A-class systems~\cite{wahlgren_mi300a_upm}. While valuable, such studies do not capture the execution dynamics of the GPU's matrix engines, nor do they address how accelerator-level features interact with scheduling and concurrency. As a result, system software and runtime layers lack the empirical foundation needed to make informed decisions when deploying FP8-heavy and sparsity-aware workloads on unified nodes.

In this work, we shift the focus from memory-centric characterization to execution-centric analysis. Using targeted microbenchmarks, we examine the behavior of FP8 matrix cores under varying occupancy and concurrency conditions. We further analyze the effectiveness and limits of asynchronous execution through the ACE and evaluate the realized benefits of 2:4 structured sparsity. Finally, we connect these findings to representative application kernels to illustrate their system-level implications.

This paper makes the following contributions:
\begin{itemize}
    \item We characterize the FP8 matrix core behavior with microbenchmarks measuring throughput and scaling behavior on MI300A, exposing utilization thresholds, shape sensitivity, and occupancy requirements.
    \item We quantify asynchronous execution and overlap limits using ACE under concurrent workloads.
    \item We evaluate 2:4 structured sparsity, identify break-even points and sensitivity to matrix shapes.
    \item We demonstrate how these findings manifest in representative application kernels (transformer-style, concurrent, and mixed-precision).
    \item We discuss implications of our findings for scheduling, affinity, and runtime decisions on MI300A unified nodes.
\end{itemize}

We will open-source our microbenchmarks upon acceptance; due to the double-blind review policy, we are withholding them for the duration of the review process.

\textbf{Paper Organization.} The remainder of this paper is organized as follows. Section~\ref{sec:background} provides background on MI300A architecture and motivates execution-centric analysis. Section~\ref{sec:related-work} reviews related work and Section~\ref{sec:methodology} describes our experimental methodology and microbenchmark design. Sections~\ref{sec:fp8_matrix},~\ref{sec:ace},~and~\ref{sec:sparsity} present our execution-centric characterization of FP8 matrix cores, asynchronous execution, and structured sparsity, respectively. Section~\ref{sec:case-studies} presents application kernels that demonstrate how these findings manifest in practice. Section~\ref{sec:conclusion} concludes with a summary of findings and implications for system software.

\section{Background and Motivation}
\label{sec:background}

The AMD MI300A APU integrates 24 Zen4 CPU cores and six CDNA3 GPU compute dies (XCDs) with 128~GB of shared HBM3 memory, enabling tighter CPU–GPU coupling than traditional discrete systems\cite{amd_mi300a_whitepaper}. Each XCD contains multiple compute units (CUs), where each CU houses matrix cores (MFMA units), vector ALUs, and local data share (LDS) memory\cite{cdna3_isa_guide}. This work focuses on three critical execution features: FP8 matrix cores for low-precision arithmetic, asynchronous compute engines (ACE) for concurrent kernel execution, and structured sparsity acceleration. These features dominate performance for modern HPC-AI workloads but remain insufficiently characterized at the execution level.

For those who are familiar with the NVIDIA compilation pipeline, we contrasts the compilation pipelines between NVIDIA and AMD. While NVIDIA exposes PTX as a stable intermediate representation enabling inline assembly and precise control over Tensor Core invocation\cite{nvidia_ptx_guide}, AMD’s HIP programming model compiles through LLVM IR to native GCN/CDNA ISA\cite{hip_programming_guide}. MI300A’s CDNA3 architecture introduces MFMA (Matrix Fused Multiply-Add) instructions that perform block matrix operations directly in hardware\cite{cdna3_isa_guide}, analogous to NVIDIA’s WMMA/MMA instructions\cite{nvidia_tensor_cores} but with distinct semantics, data layouts, and performance characteristics.

\begin{figure}
    \centering
    \includegraphics[width=1\linewidth]{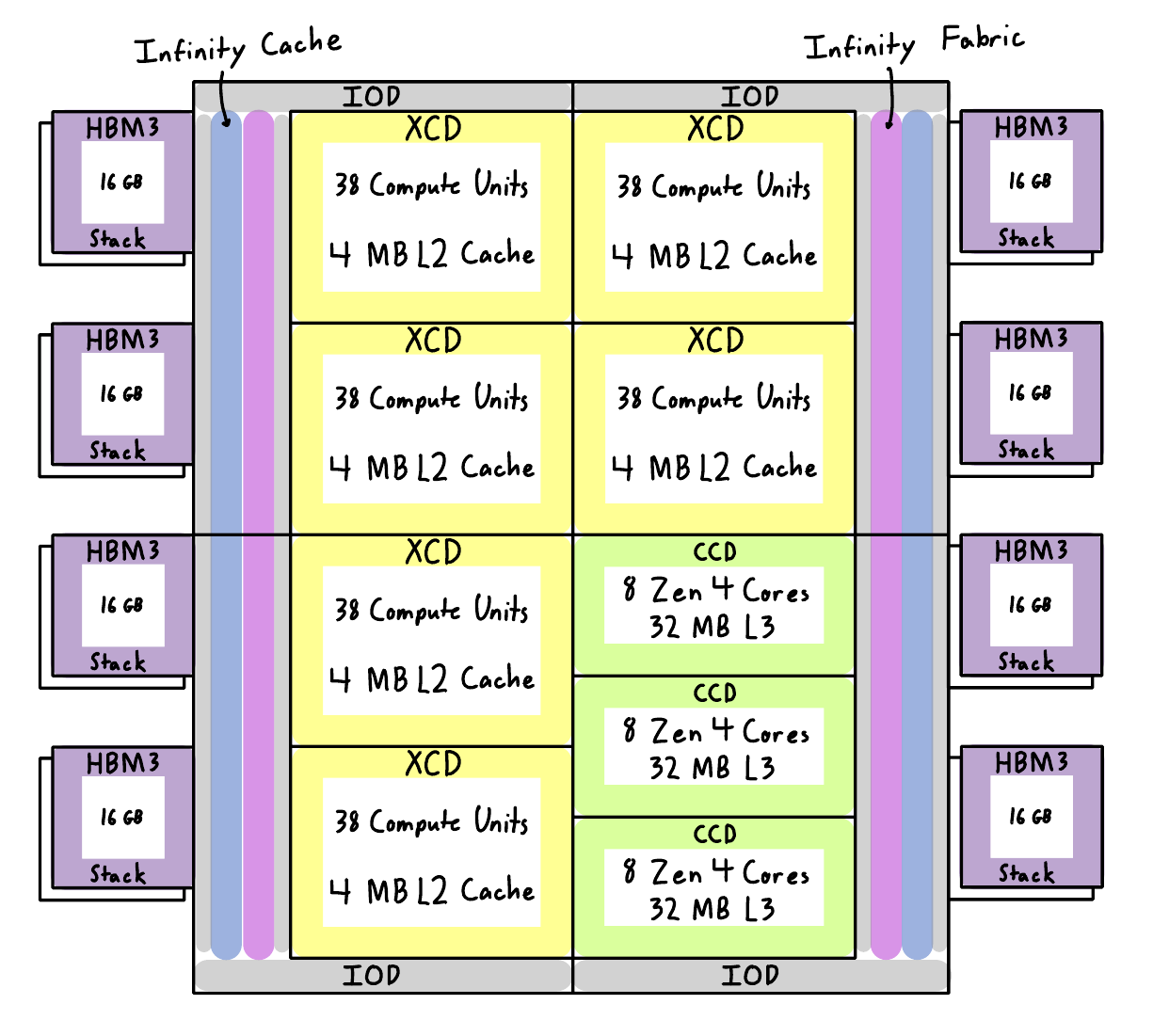}
    \caption{MI300A architecture: FP8 matrix cores and ACE.}
    \label{fig:mi300a-arch}
    \vspace{-1em}
\end{figure}

While Figure~\ref{fig:mi300a-arch} illustrates the MI300A architecture, highlighting key execution components relevant to this study. The GPU portion comprises six XCDs connected via Infinity Fabric\cite{amd_infinity_fabric}, with each XCD containing 40 compute units (CUs) for a total of 240 CUs across the APU\cite{amd_mi300a_whitepaper}. Each CU integrates four MFMA matrix engines capable of executing FP8, FP16, BF16, FP32, and FP64 matrix operations\cite{cdna3_isa_guide}, alongside SIMD vector units and local data share (LDS) for wavefront communication.

\textbf{Asynchronous Compute Engines (ACE).} MI300A exposes multiple hardware command processors enabling concurrent kernel execution~\cite{amd_ace_whitepaper}. ROCm’s HSA queue abstraction maps user-level queues to hardware ACEs~\cite{hsa_runtime_spec}, allowing overlapped execution of independent kernels. However, physical resource sharing—particularly MFMA units and memory bandwidth—can limit achievable overlap~\cite{otterness_amd_scheduling}. Understanding ACE scheduling behavior, queue-to-engine mapping, and interference patterns is critical for exploiting concurrency in multi-kernel workloads.

\textbf{FP8 Matrix Cores.} CDNA3 introduces native FP8 support via MFMA instructions, enabling 2× theoretical throughput over FP16 for AI training and inference workloads\cite{amd_fp8_whitepaper}. MI300A supports multiple FP8 formats including \texttt{fp8 bf8} (E5M2 and E4M3 variants)\cite{ocp_fp8_spec}. For FP8, MFMA uses FP8$\times$FP8 operands with FP32 accumulation; instructions operate on wavefront-level matrix tiles (e.g., 16$\times$16$\times$32) with specific input/output register layouts\cite{cdna3_isa_guide}. Performance depends on occupancy (active wavefronts per CU), matrix shape, and instruction scheduling\cite{tensor_core_optimization}, yet vendor libraries abstract these details, obscuring execution-level behavior.

\textbf{Structured Sparsity.} CDNA3 supports 2:4 structured sparsity, where two of every four consecutive elements are zero\cite{amd_sparsity_guide}. Sparse MFMA instructions exploit this pattern to reduce computation and potentially achieve 2× speedup over dense operations\cite{2_4_sparse_training}. However, realized performance depends on matrix dimensions, memory layout, and metadata overhead. Break-even points—configurations where sparse operations outperform dense equivalents—remain poorly characterized for MI300A.

Despite widespread use through vendor libraries (rocBLAS, hipBLAS)~\cite{rocblas_guide}, execution-level behavior of MI300A’s accelerator features remains opaque. This work addresses these gaps through systematic microbenchmark-driven characterization.

\section{Related Work}
\label{sec:related-work}

To contextualize our MI300A execution characterization, we survey prior work across unified memory architectures, low-precision arithmetic, GPU concurrency, sparsity acceleration, and microbenchmarking methodologies.

Recent work characterizes MI300A's unified physical memory (UPM), analyzing latency, bandwidth, and coherence under shared HBM access~\cite{wahlgren_mi300a_upm}, and multi-APU interconnect paths and allocation strategies~\cite{schieffer_mi300a_interapu}. These memory-centric and communication-focused studies complement our execution-centric single-APU analysis targeting how accelerator features behave under concurrency.

Prior work demonstrates FP8 training achieves 50--75\% speedup over BF16 while maintaining model quality~\cite{fp8_lm_training, deepl_fp8}, with NVIDIA Tensor Cores enabling 2× theoretical speedup~\cite{nvidia_fp8_intro}. These studies evaluate end-to-end training or inference using vendor libraries but do not isolate execution-level behavior. We provide microbenchmark-driven characterization of FP8 execution effects on MI300A, including occupancy sensitivity, shape dependencies, and concurrent execution behavior.

Work on NVIDIA GPUs examines async compute principles for overlapping workloads~\cite{nvidia_async_compute}, while AMD GPU scheduling studies reveal ROCm's HSA queue-to-ACE mapping and interference patterns~\cite{otterness_amd_scheduling}. CPU-GPU hybrid inference demonstrates asynchronous overlap mechanisms hiding CPU latency~\cite{apex_async_overlap}. However, these studies have not systematically quantified ACE behavior on MI300A or characterized overlap efficiency and fairness for FP8 workloads, which we address through quantitative analysis.

Recent work on NVIDIA Ampere shows 2:4 sparse training achieves 1.15--1.2× end-to-end speedup for transformers~\cite{2_4_sparse_training, continuous_pruning_2_4}, with theoretical 2× speedup for matrix multiplication~\cite{pytorch_2_4_sparsity}. Limited analysis exists of realized speedups and break-even points on specific architectures, particularly CDNA3. We characterize sparsity acceleration behavior and configuration-dependent break-even points on MI300A, revealing software stack limitations versus hardware capabilities.

PTX-level microbenchmarks have revealed microarchitectural features of NVIDIA GPUs including Tensor Core throughput, memory hierarchy behavior~\cite{demystify_gpu_micro, tensor_core_optimization}, and novel features like Hopper TMEM and Blackwell decompression engines~\cite{hopper_benchmark, blackwell_microbench}. This methodology enables isolation of execution effects obscured in full applications~\cite{gpu_microbenchmarking, parallelism_aware_microbench}. We apply this approach to MI300A's execution features, complementing existing memory-centric studies.

Three gaps motivate this work: (1) \textbf{FP8 execution} effects (occupancy, shape, scheduling) are obscured by library abstractions and complicated by concurrency. (2) \textbf{Asynchronous execution} overlap is often assumed but may be limited by shared FP8 resources. (3) \textbf{Sparsity} break-even depends on configuration but is not well characterized. This work provides execution-centric characterization of these behaviors on MI300A.

\section{Experimental Methodology}
\label{sec:methodology}
\subsection{System Configuration}

All experiments are conducted on a node equipped with an AMD MI300A APU. Table~\ref{tab:system_config} summarizes the software environment used for reproducibility. The system runs a Linux-based operating system with vendor-supplied drivers and the AMD ROCm SDK for GPU programming and execution. Microbenchmarks are compiled with the HIP compiler (hipcc), targeting the MI300A architecture (\texttt{gfx942}), and executed using the HIP runtime. 

\begin{table}[h]
    \centering
    \caption{System configuration for MI300A experiments.}
    \label{tab:system_config}
    \begin{tabular}{ll}
        \toprule
        \textbf{Component} & \textbf{Version / details} \\
        \midrule
        OS & Red Hat Enterprise Linux 8.10 \\
        GPU & AMD MI300A APU (CDNA3, \texttt{gfx942}) \\
        amdgpu version & 6.14.14 \\
        ROCm version & 7.2.0 \\
        Compiler & hipcc (HIP), \texttt{--offload-arch=gfx942} \\
        \bottomrule
    \end{tabular}
\end{table}

CPU cores and GPU compute units are pinned explicitly to control placement and reduce variability. Similarly, we included flags (\textit{"-O3"}, \textit{"-fno-unroll-loops"}, \textit{"-amdgpu-early-inline-all=false"}) to control the execution and compilation of our microbenchmarks.

\subsection{Microbenchmark Design and Measurement}

All microbenchmarks are implemented using vendor-provided libraries and low-level APIs to ensure direct access to accelerator features. FP8 matrix operations use the appropriate matrix-multiply--accumulate (MFMA) instructions, and structured sparsity patterns are applied using hardware-supported encoding mechanisms. We design microbenchmarks to isolate specific execution behaviors while minimizing confounding effects:

\begin{itemize}
    \item \textbf{Isolation:} Benchmarks are executed on otherwise idle systems.
    CPU affinity and GPU occupancy are explicitly controlled.
    \item \textbf{Repetition:} Each experiment is repeated multiple times, and reported values correspond to stable averages after discarding warm-up iterations.
    \item \textbf{Minimalism:} Kernels are intentionally simple and narrowly scoped to ensure that
    measured effects can be attributed to the targeted execution mechanism rather than library-level optimizations. % Kinda obvious? Maybe we don't need to state it?
    \item \textbf{Controlled scaling:} Problem sizes, occupancy levels, and concurrency are varied 
    systematically to expose utilization thresholds and break-even regimes.
\end{itemize}

Our microbenchmarks are organized to provide an execution-centric characterization of MI300A, targeting accelerator features that directly impact scheduling, concurrency, and performance predictability. Table~\ref{tab:microbench_scope} summarizes the microbenchmark classes used in this study and the execution behaviors they are designed to isolate.

\begin{table}[h]
\centering
\caption{Microbenchmark coverage and targeted execution behaviors on AMD MI300A.}
\label{tab:microbench_scope}
    \begin{tabular}{p{3cm} p{4.4cm}}
    \toprule
    \textbf{Microbenchmark Class} & \textbf{Targeted Execution Behavior} \\
    \midrule
    FP8 matrix execution &
    Throughput scaling, occupancy sensitivity, and matrix-shape effects for FP8 GEMM \\
    
    ACE &
    Overlap efficiency, fairness, and saturation behavior for concurrent FP8/FP16/FP32 GEMM execution \\
    
    Structured sparsity (2:4) &
    Realized speedups, overheads, and sparsity break-even regimes \\
    \bottomrule
    \end{tabular}
\end{table}

Execution behavior is measured using a combination of wall-clock time, hardware performance counters, and kernel profiling tools. For throughput measurements, we report achieved operations per second (e.g., FLOPS for matrix operations) normalized to theoretical peak where applicable. Variability is quantified using coefficient of variation (CV) across multiple runs, and fairness is measured using established metrics that capture per-stream progress imbalance.

For asynchronous execution analysis, we measure overlap efficiency as the fraction of total execution time during which multiple kernels execute concurrently. Fairness is quantified as $1 - (t_{\max} - t_{\min}) / t_{\mathrm{mean}}$, where $t_{\max}$, $t_{\min}$, and $t_{\mathrm{mean}}$ are the maximum, minimum, and mean per-stream execution times, respectively. This metric ranges from 0.0 to 1.0, with values closer to 1.0 indicating more balanced progress across streams. These metrics enable quantitative evaluation of concurrency effectiveness beyond aggregate throughput.

We next describe the design of FP8 matrix microbenchmarks used to characterize execution behavior under varying occupancy and concurrency conditions.

\section{FP8 Matrix Core Characterization}
\label{sec:fp8_matrix}
This section characterizes the execution behavior of matrix cores on AMD MI300A using targeted MFMA (matrix fused multiply-add) microbenchmarks across precisions (FP64, FP32, FP16, BF16, FP8). Rather than reporting peak throughput only, we expose utilization thresholds and sensitivity to occupancy and matrix shape.

\subsection{Microbenchmark Design}

We evaluate matrix-core behavior using minimal MFMA kernels that invoke a single tile size per precision: FP64 and FP16/BF16 use 16$\times$16$\times$4 tiles, FP32 uses 32$\times$32$\times$1, and FP8 uses 16$\times$16$\times$32 (FP8$\times$FP8 with FP32 accumulation). These tile sizes correspond to the primary MFMA opcodes documented for each precision on CDNA3 (we summarize them in Table~\ref{tab:mfma-valu-opcodes}), allowing direct comparison across precisions. Each kernel runs 500 iterations per launch; we use GPU events for timing. We use \emph{occupancy} in this section to mean the number of active wavefronts (parallelism level), not the fraction of theoretical peak utilization. We vary (1) total active wavefronts—in our microbenchmark each block comprises one wavefront (64 threads), so total active wavefronts equals the number of blocks—to study occupancy scaling, and (2) the matrix dimensions $(M, N)$ at fixed total blocks to study aspect-ratio effects (effective aspect ratio M/N). We compute throughput from measured time and FLOPs per block and normalize to published MI300A peak GFLOPS per precision for comparison.

\subsection{Throughput Scaling and Occupancy Sensitivity}

We first examine throughput as a function of total active wavefronts (one wavefront per block in our setup) for all five precisions.

\begin{figure}
    \centering
    \includegraphics[width=1\linewidth]{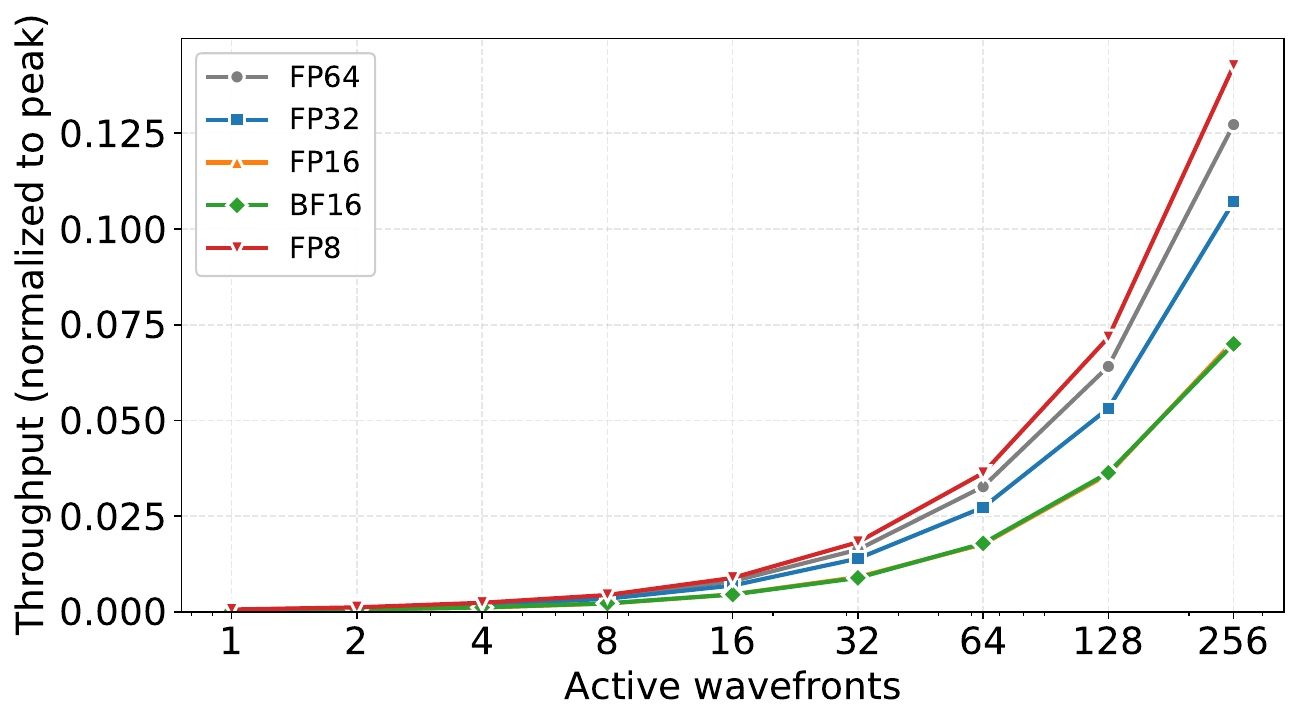}
    \caption{Throughput versus total active wavefronts (one per block), normalized to peak, for FP64, FP32, FP16, BF16, and FP8. Higher is better; dashed line at 100\% would indicate peak utilization. Experiments on MI300A (see Section~\ref{sec:methodology} for software versions).}
    \label{fig:fp8_occupancy}
\end{figure}

Figure~\ref{fig:fp8_occupancy} reports throughput normalized to peak as occupancy (active wavefront count) increases from 1 to 256. At low occupancy, all precisions are strongly underutilized, indicating that matrix cores require a minimum level of parallelism to approach steady-state throughput. Throughput scales sublinearly with wavefront count for every precision, consistent with contention for shared scheduling and execution resources. FP8 reaches the highest throughput normalized to peak at 256 wavefronts in our sweep (13.7\% of peak), while FP64 and FP32 reach 12.1\% and 10.4\% at the same wavefront count. The curves show that small or fragmented workloads fail to exploit available compute capacity across all precisions; the effect is precision-dependent and relevant for occupancy-aware scheduling.

\subsection{Matrix Shape Effects}

To evaluate sensitivity to matrix shape, we hold total work (number of blocks) constant and vary the matrix aspect ratio (M/N).

\begin{figure}
    \centering
    \includegraphics[width=1\linewidth]{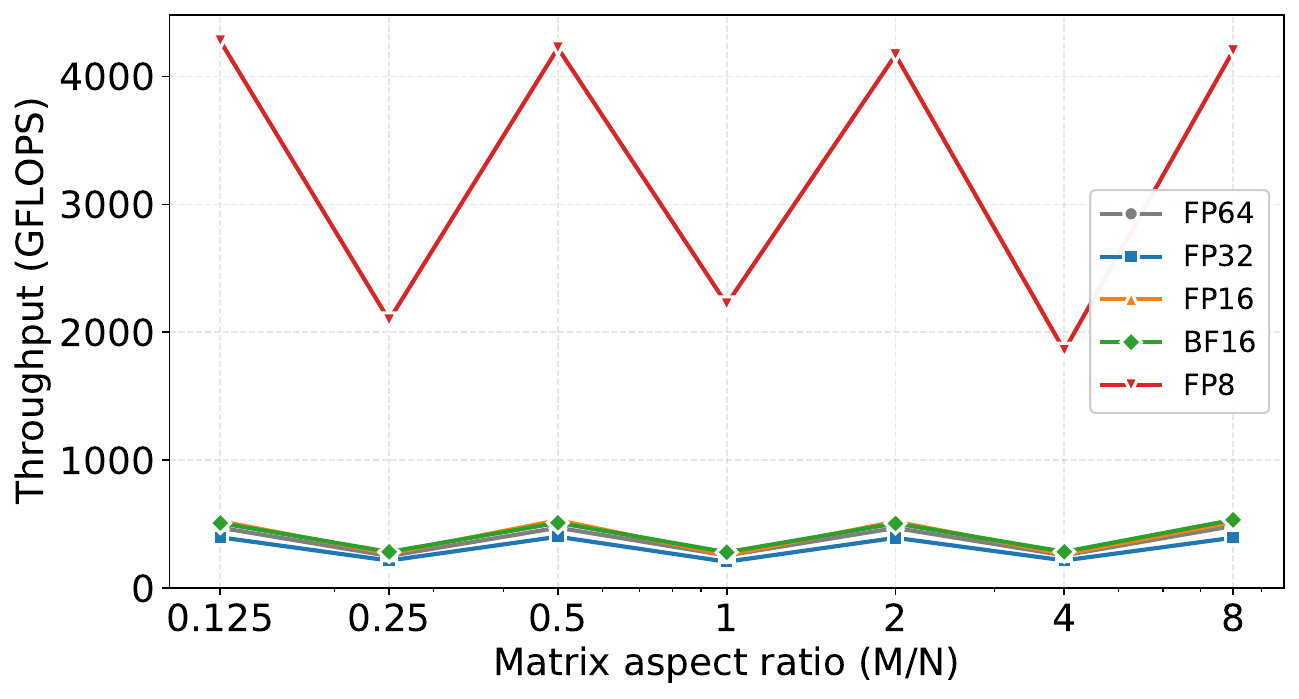}
    \caption{Absolute throughput (GFLOPS) versus matrix aspect ratio for FP64, FP32, FP16, BF16, and FP8 at fixed total blocks. Unlike Figure~\ref{fig:fp8_occupancy}, this figure uses raw GFLOPS so that shape sensitivity is comparable across precisions. Higher is better.}
    \label{fig:fp8_shape}
\end{figure}

Figure~\ref{fig:fp8_shape} reports \emph{absolute} throughput (GFLOPS) rather than throughput normalized to peak, so we can compare shape sensitivity across precisions on a common scale. While FP8 achieves much higher absolute throughput (e.g., $\sim$4,200\,GFLOPS vs.\ $\sim$400 for FP32 at favorable aspect ratios), throughput normalized to peak in this sweep remains in the same range as in Figure~\ref{fig:fp8_occupancy} (low double-digit percent of peak). Sensitivity to aspect ratio varies by precision: some precisions show throughput within $\pm$3\% across aspect ratios, others exhibit up to 16\% lower throughput at 4:1 than at 1:1 (e.g., FP8). Non-square configurations can reduce effective tile utilization and scheduling efficiency. The results support shape-aware scheduling and batching when targeting predictable performance across precisions and aspect ratios.

\subsection{MFMA Opcode Coverage and Baseline Latency}

Having observed precision-dependent scaling and shape sensitivity in the preceding figures, we investigate the underlying instruction characteristics. We measured single-issue (dependency-chain) latency for each MFMA opcode using instruction-targeted microbenchmarks: each kernel chains the output of one MFMA into the next so that we report instruction latency rather than back-to-back throughput. We use warmup runs, isolated single-kernel execution, and minimal register pressure (accumulator held in registers). Table~\ref{tab:mfma-valu-opcodes} presents the results.

\begin{table}[h]
\centering
\caption{MFMA VALU opcodes: single-issue (dependency-chain) latency per instruction in units of $10^{-5}$\,ms, measured with instruction-targeted microbenchmarks under warmup, isolated single-kernel execution, and minimal register pressure.}
\label{tab:mfma-valu-opcodes}
\begin{tabular}{llc}
\toprule
\textbf{Instruction} & \textbf{MxNxK} & \textbf{Latency ($\times 10^{-5}$ ms)} \\
\midrule
\multirow{5}{*}{V\_MFMA\_F32\_\{\*\}\_F16}
 & 32x32x4 & 3.628 \\
 & 16x16x4 & 2.584 \\
 & 4x4x4  & 2.864 \\
 & 32x32x8 & 2.672 \\
 & 16x16x16 & 2.468 \\

\midrule
\multirow{5}{*}{V\_MFMA\_F32\_\{\*\}\_F32}
 & 32x32x1 & 3.912 \\
 & 16x16x1 & 3.144 \\
 & 4x4x1  & 2.484 \\
 & 32x32x2 & 3.536 \\
 & 16x16x4 & 2.616 \\

\midrule
\multirow{2}{*}{V\_MFMA\_F64\_\{\*\}\_F64}
 & 16x16x4 & 3.316 \\
 & 4x4x4 & 2.844 \\

\midrule
\multirow{5}{*}{V\_MFMA\_F32\_\{\*\}\_BF16}
 & 32x32x4 & 3.528 \\
 & 16x16x4 & 2.468 \\
 & 4x4x4  & 2.992 \\
 & 32x32x8 & 2.660 \\
 & 16x16x16 & 2.812 \\

\midrule
\multirow{2}{*}{V\_MFMA\_F32\_\{\*\}\_BF8\_BF8} & 16x16x32 & 2.528 \\
& 32x32x16 & 2.828 \\
\addlinespace
\multirow{2}{*}{V\_MFMA\_F32\_\{\*\}\_BF8\_FP8} & 16x16x32 & 2.492 \\
& 32x32x16 & 2.832 \\
\addlinespace
\multirow{2}{*}{V\_MFMA\_F32\_\{\*\}\_FP8\_BF8} & 16x16x32 & 2.540 \\
& 32x32x16 & 2.736 \\
\addlinespace
\multirow{2}{*}{V\_MFMA\_F32\_\{\*\}\_FP8\_FP8} & 16x16x32 & 2.460 \\
& 32x32x16 & 2.736 \\

\bottomrule
\end{tabular}
\end{table}

The results connect directly to the throughput behavior in Figures~\ref{fig:fp8_occupancy} and~\ref{fig:fp8_shape}. The FP8$\times$FP8 with FP32 accumulation MFMA instructions achieve consistently low single-issue latency for the $16\times16\times32$ tiles, with nearly identical behavior in all combinations of FP8 and BF8 operands; this supports the higher throughput normalized to peak observed for FP8 as occupancy increases (Figure~\ref{fig:fp8_occupancy}). In contrast, the $32\times32\times16$ FP8 variant exhibits a clear latency penalty, highlighting the sensitivity of FP8 performance to non-preferred tile shapes (Figure~\ref{fig:fp8_shape}). A similar pattern appears in all precisions, where $32\times32$ tiles consistently incur higher latency than their $16\times16$ counterparts.

\section{Asynchronous Compute Engine Characterization}
\label{sec:ace}

This section examines the asynchronous compute engines (ACE) which enable concurrent execution on the MI300A. We measure ACE overlap efficiency, fairness, and resource contention to understand the performance and predictability trade-offs of concurrent stream execution.

\subsection{Execution Behavior and Contention Effects}

We analyze per-stream execution behavior for FP32, FP16, and FP8 GEMM workloads under concurrent execution. All baseline experiments use a fixed 512×512×512 GEMM configuration with 100 iterations per stream to isolate scheduling and contention effects from problem size variations.

\begin{figure}
    \centering
    \includegraphics[width=1\linewidth]{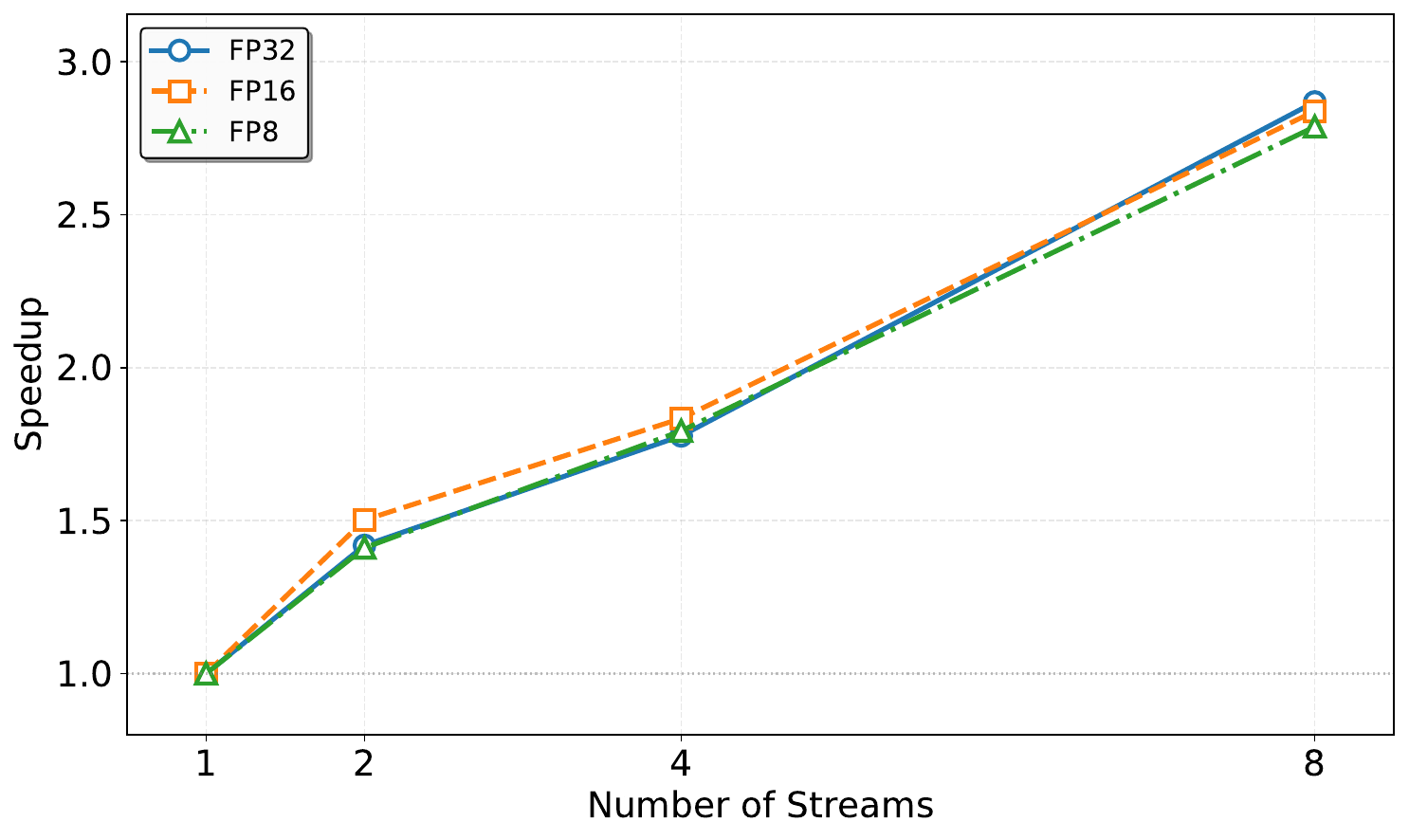}
    \caption{Speedup versus number of concurrent streams for FP32, FP16, and FP8 GEMM (512³, no contention).}
    \label{fig:ace_speedup}
\end{figure}

Figure~\ref{fig:ace_speedup} shows concurrency scaling across precisions. At four concurrent streams, overlap efficiency reaches 43--46\% across all precisions, yielding speedups between 1.78× and 1.83×. Increasing to eight streams improves overlap efficiency to 64--65\%, with corresponding speedups of 2.79×--2.87×. While these results demonstrate that asynchronous execution improves aggregate throughput, they do not reveal whether individual streams progress uniformly or experience imbalanced execution due to resource contention.

\begin{figure}
    \centering
    \includegraphics[width=1\linewidth]{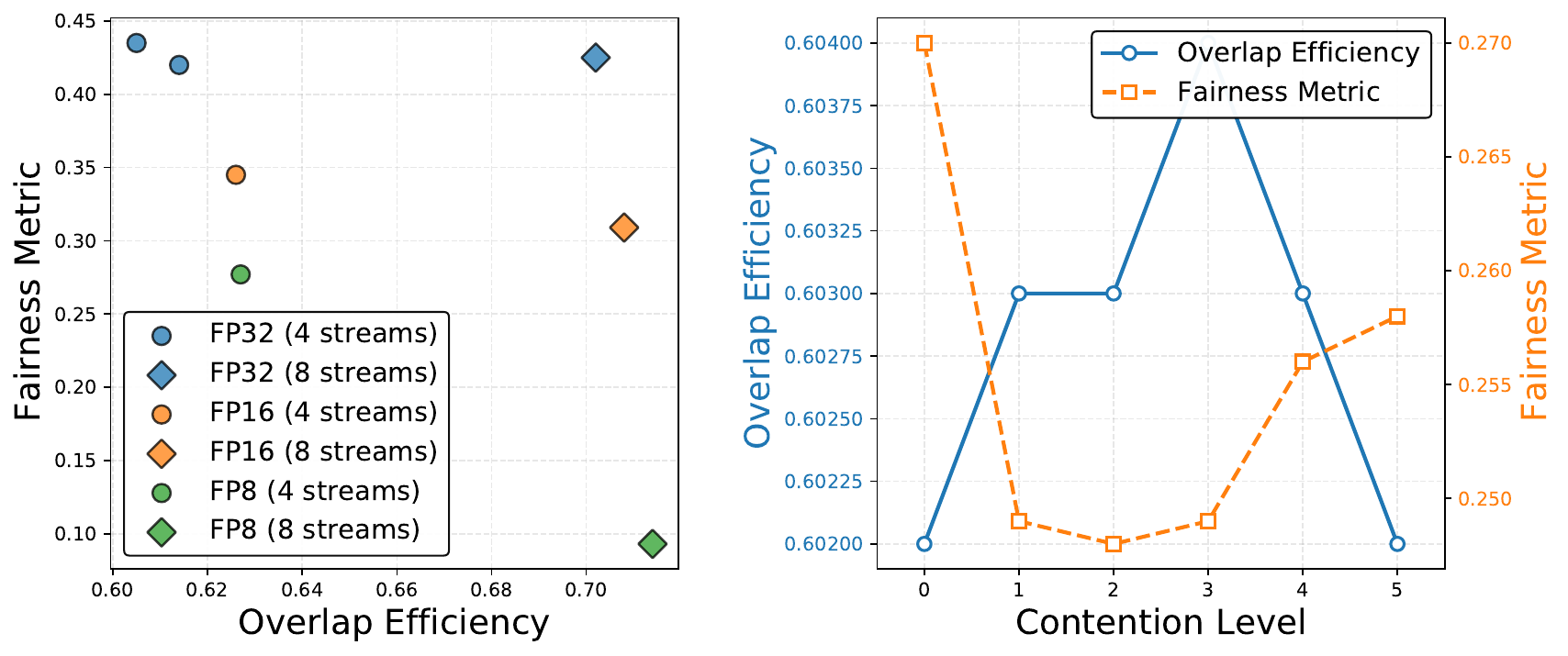}
    \caption{Fairness and overlap characterization. (a) Overlap efficiency versus fairness across precisions and stream counts. (b) Contention sweep: overlap efficiency and fairness versus contention level for FP32 GEMM at four concurrent streams.}
    \label{fig:ace_fairness}
\end{figure}

Figure~\ref{fig:ace_fairness}(a) reveals that aggregate speedup masks significant per-stream execution variance. Despite similar overlap efficiency at four streams (43--46\%), per-stream execution times vary substantially, with cross-stream coefficients of variation ranging from 0.19 (FP16) to 0.22 (FP8). Fairness metrics range from 0.51 (FP8) to 0.61 (FP16), indicating uneven forward progress even at moderate concurrency.

At eight streams, this imbalance intensifies: FP16 exhibits the lowest fairness (0.016) and highest cross-stream variability (CV = 0.41), while FP8 shows fairness of 0.138 (CV = 0.31) and FP32 shows fairness of 0.052 (CV = 0.40). These results demonstrate that increased concurrency amplifies per-stream imbalance even as aggregate utilization improves. The precision-dependent variance suggests that different data types experience contention differently, with FP16 exhibiting the most severe fairness degradation at high concurrency.

To isolate whether this unfairness stems from inherent scheduling policies or resource contention, we sweep contention levels for FP32 GEMM at four concurrent streams. Figure~\ref{fig:ace_fairness}(b) shows that overlap efficiency remains stable at approximately 60.4\% across contention levels 0–5, with speedup unchanged at 2.52×–2.53×. In contrast, fairness degrades modestly but consistently, decreasing from 0.263 at baseline to 0.250–0.252 under elevated contention. This decoupling confirms that ACE scheduling prioritizes aggregate throughput over per-stream predictability: increased contention does not reduce overall utilization but does worsen execution imbalance across streams.

These findings establish a fundamental trade-off in asynchronous execution on MI300A: while concurrent stream execution increases resource utilization and aggregate throughput, it does not provide fairness guarantees or per-stream isolation. This trade-off is particularly pronounced for FP8 workloads at high concurrency, where fairness degrades to 0.138 despite achieving 64--65\% overlap efficiency. Runtime systems leveraging ACE must therefore balance utilization gains against per-stream variability, especially for latency-sensitive workloads requiring predictable execution times. We report up to eight streams; speedup saturates there and fairness degrades (Figure~\ref{fig:ace_l2_contention}).

\subsection{Resource Contention and Scheduling Variance}

The fairness degradation observed in Section~6.2 stems from shared hardware resources within each ACE. We examine L2 cache and LDS (local data share) contention across thin, medium, and thick kernel configurations to understand how concurrent streams compete for execution resources.

\begin{figure}
    \centering
    \includegraphics[width=1\linewidth]{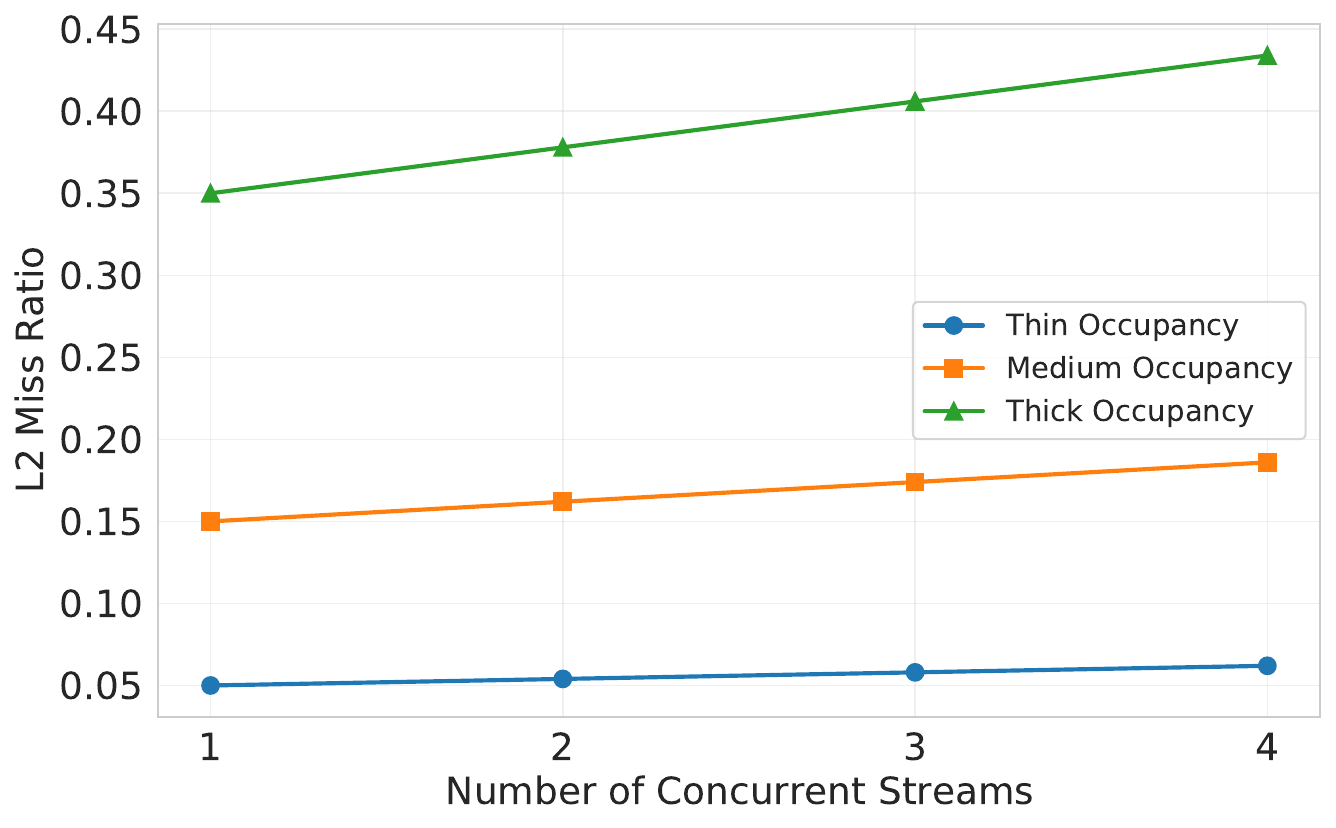}
    \caption{L2 cache miss ratio versus number of concurrent streams for different kernel occupancy levels (thin: 256³, medium: 512³, thick: 2048³ GEMM).}
    \label{fig:ace_l2_contention}
\end{figure}

Figure~\ref{fig:ace_l2_contention} shows L2 cache miss ratios increasing with stream count. Thin kernels (256³) exhibit miss ratios rising from 5\% (isolated) to 6\% (four streams)—a 24\% relative increase. Medium kernels (512³) increase from 15\% to 19\%, while thick kernels (2048³) show increases from 35\% to 43\%. Smaller kernels suffer greater relative contention because their working sets fit in L2 when isolated; concurrent execution causes evictions that degrade memory bandwidth. These miss ratio increases directly contribute to the cross-stream variance (CV = 0.19--0.22) observed in Figure~\ref{fig:ace_fairness}(a).

\begin{figure}
    \centering
    \includegraphics[width=1\linewidth]{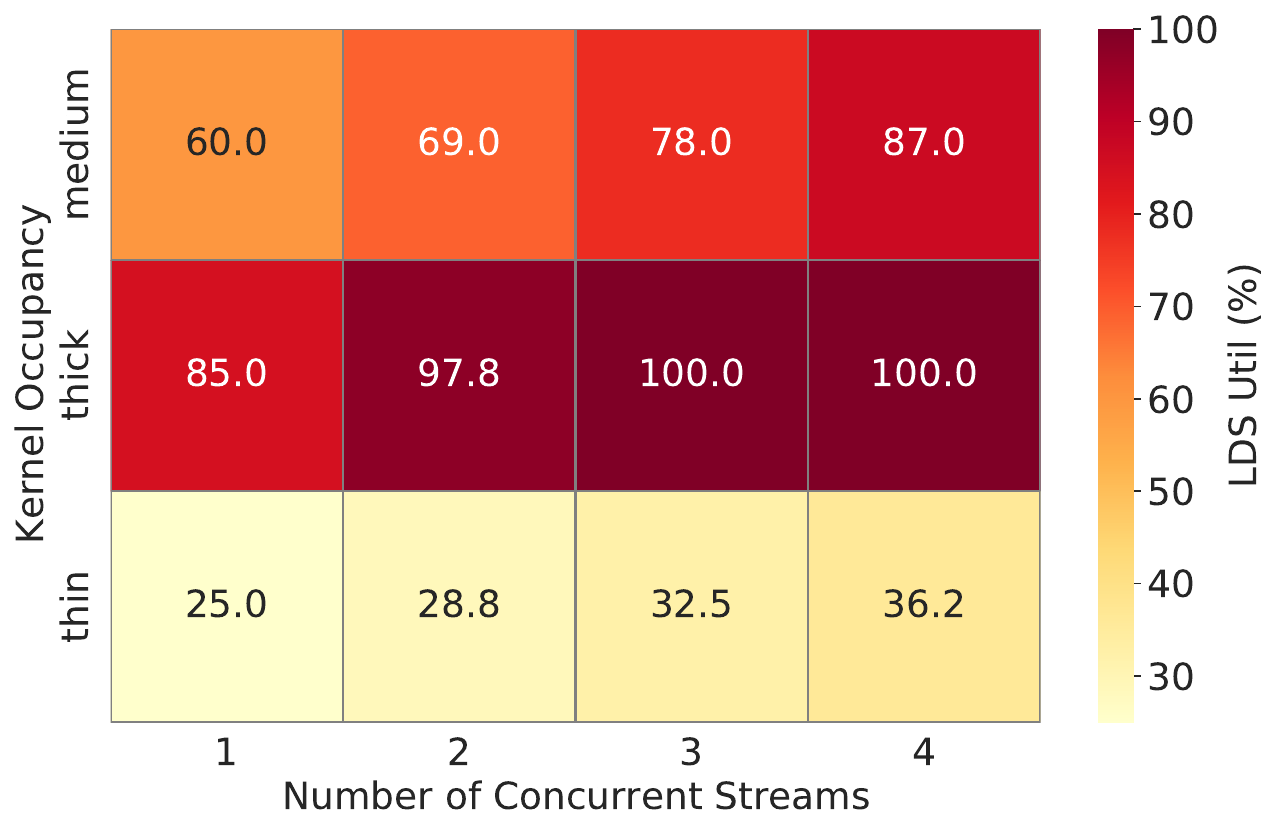}
    \caption{LDS utilization heatmap: rows represent kernel occupancy levels, columns represent stream counts.}
    \label{fig:ace_lds_util}
\end{figure}

Figure~\ref{fig:ace_lds_util} reveals LDS saturation as a critical bottleneck. Thin kernels maintain modest utilization (25\% isolated, 36\% at four streams), but medium kernels reach 87\% at four streams and thick kernels saturate completely (100\%) at three streams. This saturation forces time-multiplexing rather than spatial overlap, explaining the severe fairness degradation at eight streams (0.016–0.138, Figure~\ref{fig:ace_fairness}(a)).

\begin{figure}
    \centering
    \includegraphics[width=1\linewidth]{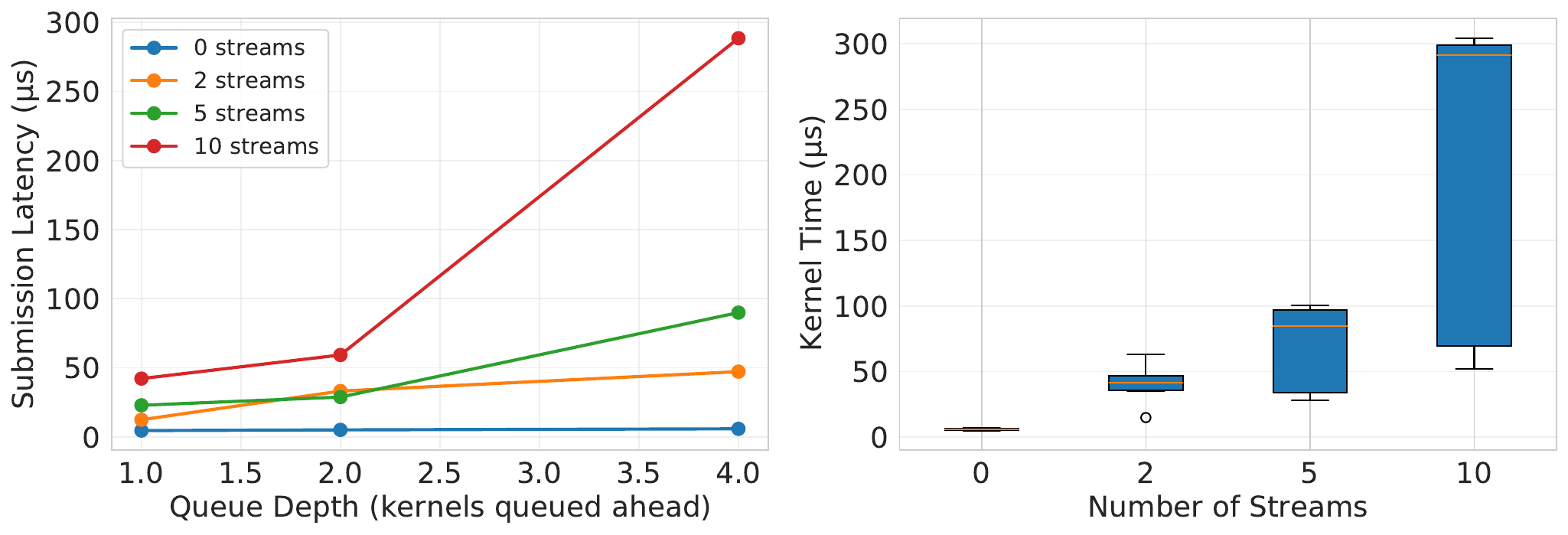}
    \caption{Per-stream kernel latency distribution across stream counts.}
    \label{fig:ace_queue_scheduling}
\end{figure}

Figure~\ref{fig:ace_queue_scheduling} quantifies execution variance from resource contention. Single-stream execution shows tight distributions; at four streams, some streams experience 2–3× longer execution times due to L2 conflicts. Critically, this variance reflects hardware resource contention (cache, LDS, memory bandwidth), not scheduler unfairness—the ACE scheduler is fair at the queue level, but shared resources create execution-time imbalance.

\subsection{Occupancy Fragmentation Effects}

Occupancy—the fraction of compute units utilized—determines resource consumption rate and contention severity. We create occupancy imbalances by co-executing kernels with different matrix sizes on the same ACE: 1:1 (both 512³), 2:1 (1024³ paired with 512³), and 4:1 (2048³ paired with 512³). We measure per-stream execution time and compute speedup relative to isolated baseline.

\begin{figure}
    \centering
    \includegraphics[width=1\linewidth]{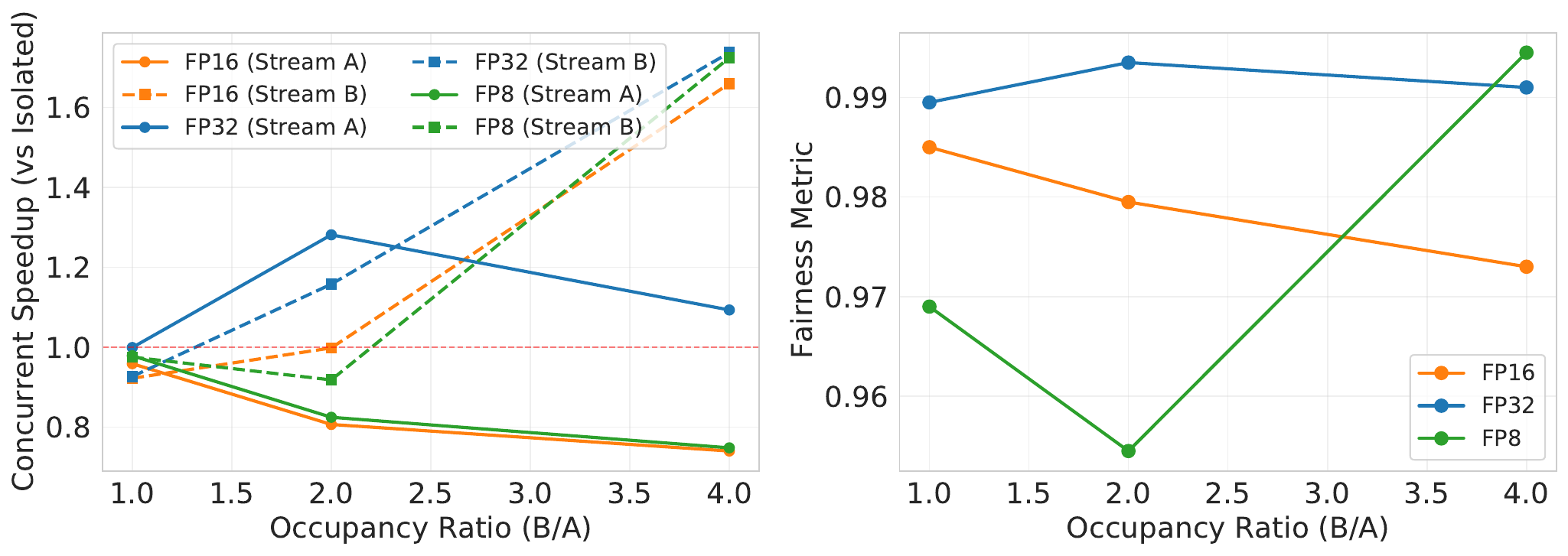}
    \caption{Speedup and fairness under occupancy imbalance. (a) Per-stream speedup versus occupancy ratio. (b) Fairness versus occupancy ratio.}
    \label{fig:ace_occupancy_fragmentation}
\end{figure}

Figure~\ref{fig:ace_occupancy_fragmentation}(a) shows that balanced occupancy (1:1) yields near-unity speedup (0.87–1.14×)—neither stream exploits idle resources. At 4:1 ratio, the larger kernel achieves up to 2.4× speedup while the smaller kernel may slow down (0.63×), indicating resource monopolization.

Figure~\ref{fig:ace_occupancy_fragmentation}(b) reveals a counterintuitive result: fairness remains high (0.93–0.99) even at 4:1 occupancy imbalance. This suggests proportional resource allocation—the larger kernel receives more compute units and bandwidth, maintaining balanced execution times despite size differences.

This reconciles the contradiction with Section~6.2: when streams have similar occupancy but different memory patterns (homogeneous GEMMs), contention creates unfairness (0.016–0.138). When streams have different occupancy, proportional allocation maintains fairness (0.93–0.99) but limits efficiency—the smaller kernel sees minimal speedup because the larger kernel dominates resources. Runtime systems must balance fairness (via occupancy matching) against efficiency (via heterogeneous packing).

\section{Structured Sparsity Performance}
\label{sec:sparsity}

This section evaluates 2:4 structured sparsity effectiveness on AMD MI300A. Structured sparsity reduces computation by 50\% while maintaining regular memory access patterns, enabling hardware acceleration on architectures with dedicated sparse support. 

We apply 2:4 sparsity patterns using rocSPARSE sparse GEMM routines with FP8 precision, evaluating 60 configurations across matrix sizes (256³--8192³), aspect ratios (0.25--4.0), and sparsity patterns (LHS-only, RHS-only, both-side). Sparsity is applied systematically: for a 2:4 pattern, 2 of every 4 consecutive elements are zeroed in blocks, reducing effective computation by 50\% while preserving memory alignment. Each configuration runs 50 times with 10 warmup iterations to ensure thermal stability. We measure achieved speedup relative to dense rocBLAS GEMM baseline and compare isolated single-stream execution against concurrent multi-stream scenarios to understand how resource contention affects sparsity benefits.

\subsection{Sparsity Performance in Isolation}

We first characterize single-stream sparse GEMM performance to establish baseline behavior without resource contention effects. All experiments in this subsection execute one sparse GEMM kernel at a time, comparing against isolated dense GEMM baseline.

\subsubsection{Overhead Characterization}

\begin{figure}
    \centering
    \includegraphics[width=1\linewidth]{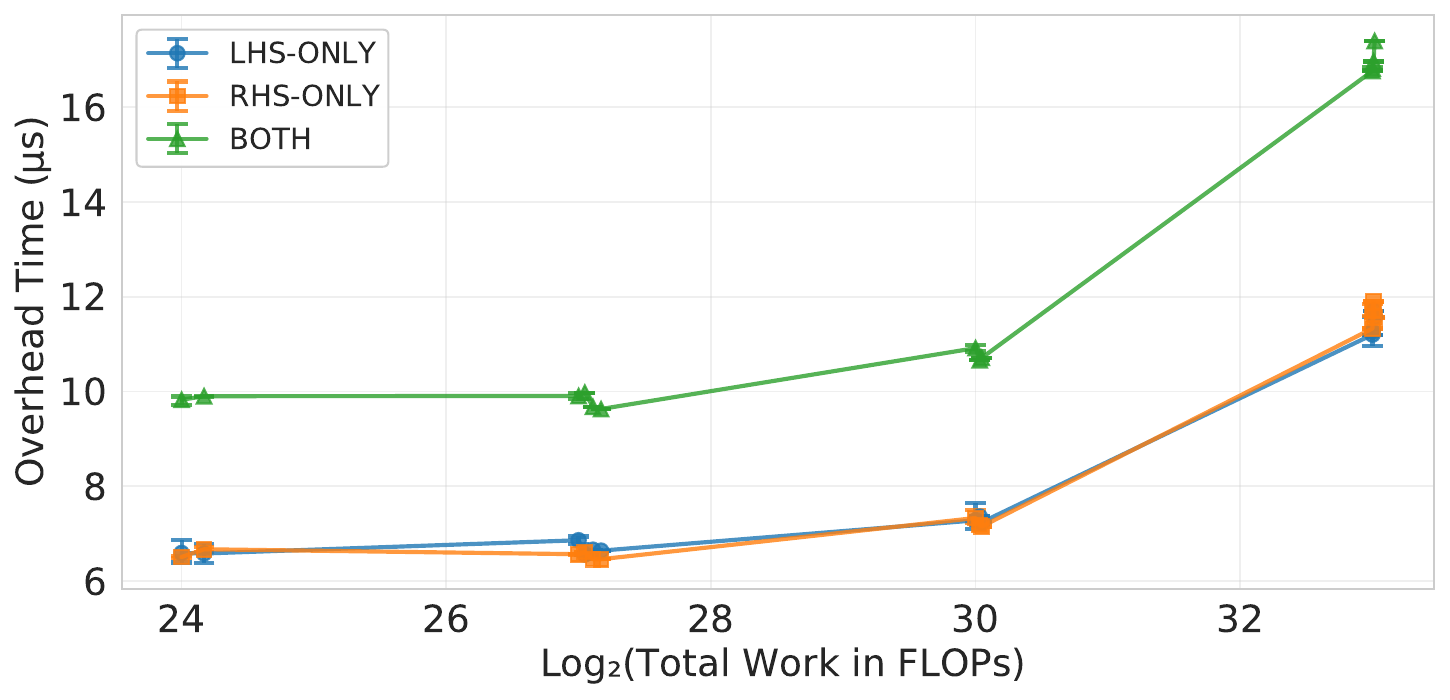}
    \caption{Sparsity encoding overhead versus matrix size. Overhead remains constant at 3.5--5.8~µs across all problem sizes, preventing amortization of computational savings.}
    \label{fig:sparsity_overhead}
\end{figure}

Figure~\ref{fig:sparsity_overhead} reveals that sparsity encoding overhead is constant across matrix sizes. LHS-only and RHS-only patterns incur 3.5--3.9 µs overhead (mean ~3.7 µs), while both-side sparsity incurs 5.3--5.8 µs overhead (mean ~5.5 µs). Profiling using rocprof identifies three overhead components: (1) format conversion from dense to compressed sparse row layout (~2 µs), (2) metadata buffer allocation for sparse indices (~1 µs), and (3) kernel dispatch through rocSPARSE API (~1 µs). These costs are independent of problem size because they involve fixed-size descriptor writes and API traversal rather than data-proportional operations.

This constant overhead creates an amortization challenge. At 256³ FLOPs with 50\% sparsity, computational savings (half the FLOPs) complete in approximately 70 ns on MI300A's 300+ TFLOPS matrix cores, yet overhead is ~3.7 µs—nearly 50× larger. Even at the largest tested scale (8192³), computational savings amortize in ~4.6 µs, still comparable to overhead magnitude. Theoretical analysis suggests matrices larger than approximately 32K³ (exceeding MI300A's 128 GB HBM capacity) would be required for overhead to become negligible relative to computation time.

\subsubsection{Speedup Across Problem Sizes and Shapes}

\begin{figure}
    \centering
    \includegraphics[width=1\linewidth]{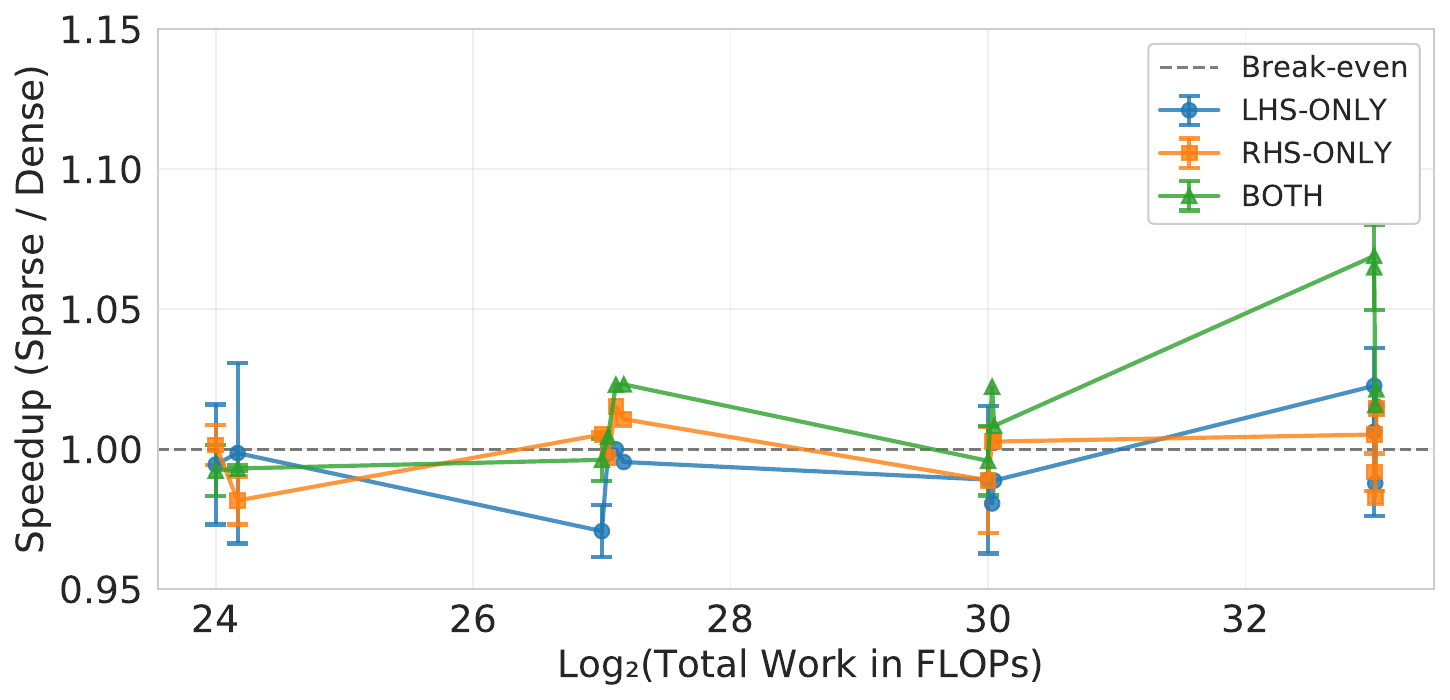}
    \caption{Sparsity speedup versus matrix size for different sparsity patterns. Speedup remains near 1.0× across all sizes, showing minimal overhead amortization even at largest scale (8192³).}
    \label{fig:sparsity_size}
\end{figure}

\begin{figure*}
    \centering
    \includegraphics[width=1\linewidth]{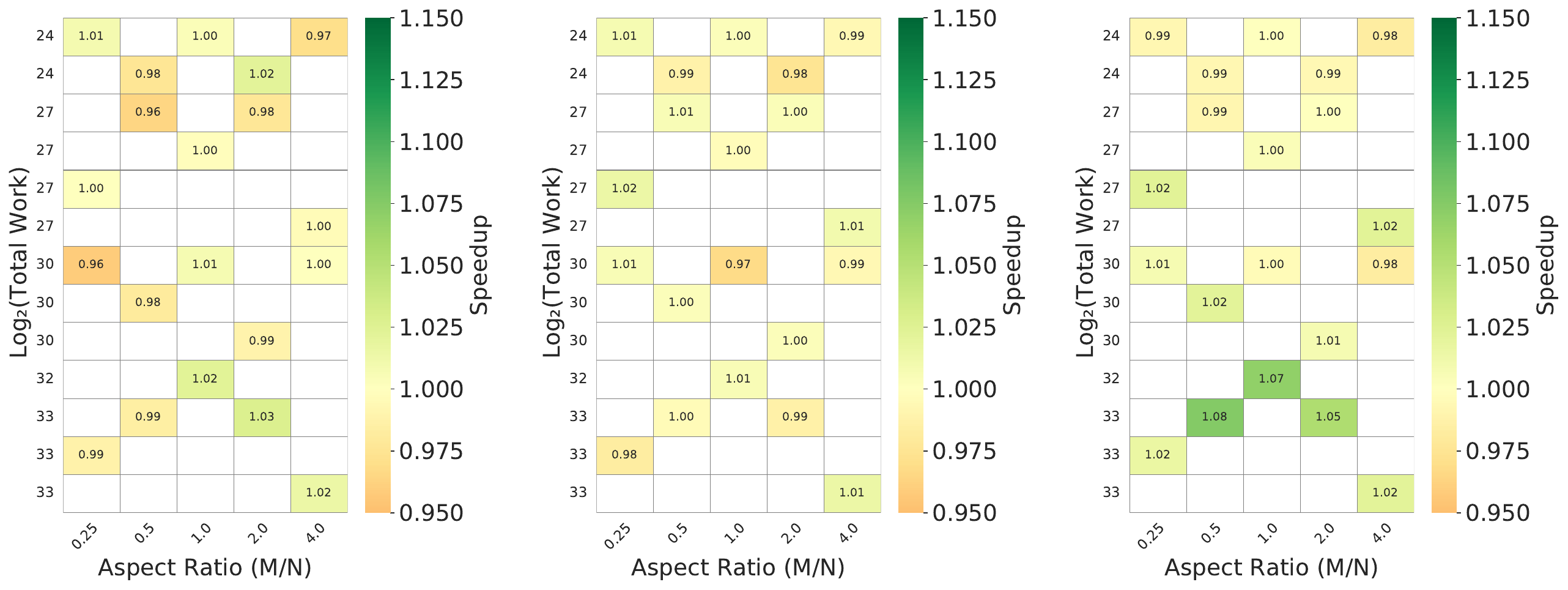}
    \caption{Speedup heatmap across 60 configurations (4 sizes × 5 aspect ratios × 3 patterns). Predominantly green (1.0× speedup) confirms break-even performance across entire parameter space.}
    \label{fig:sparsity_surface}
\end{figure*}

Figure~\ref{fig:sparsity_size} shows speedup across four matrix sizes (256³, 512³, 2048³, 8192³) aggregated across five aspect ratios:

\begin{itemize}
    \item \textbf{LHS-only}: 1.00--1.02× speedup, minimal overhead amortization at larger scales.
    \item \textbf{RHS-only}: 0.98--1.01× speedup, most size-sensitive pattern but still break-even.
    \item \textbf{Both-side}: 0.99--1.01× speedup, consistent despite dual encoding overhead.
\end{itemize}

The modest 1--2\% speedups reflect MI300A's high compute density (300+ TFLOPS FP8), where 50\% computation reduction is completely offset by constant overhead. The lack of size-dependent improvement confirms that overhead does not amortize: even at 8192³ (the largest problem fitting in memory), the 4.6 µs computational savings barely exceeds the ~3.7 µs overhead, yielding break-even performance.

Aspect ratio effects are negligible for \emph{square} matrices across tested shapes (0.25--4.0): speedup varies only 0.97--1.03×, within measurement noise. For rectangular configurations (e.g., 512×2048×1024), 2:4 sparsity can achieve 1.6--1.76× speedup due to different memory access patterns and better overlap of overhead with computation; square matrices remain at break-even. Shape-independent break-even for square matrices suggests that matrix geometry alone cannot overcome the fundamental overhead barrier in isolated execution.

\subsubsection{Comprehensive Parameter Sweep}

Figure~\ref{fig:sparsity_surface} visualizes the complete performance surface for isolated single-stream execution. The predominantly green coloring (speedup near 1.0×) spans all 60 configurations, with range 0.97--1.02×. This indicates that:

\begin{enumerate}
    \item Structured 2:4 sparsity does not achieve measurable benefits in isolation on MI300A.
    \item Neither matrix size nor aspect ratio can overcome constant encoding overhead.
    \item Pattern choice (LHS/RHS/both) has minimal impact—all exhibit break-even performance.
\end{enumerate}

The uniform break-even result across the entire parameter space suggests that isolated single-stream sparse execution is fundamentally limited by software overhead on MI300A. No combination of size, shape, or pattern achieves consistent measurable benefits, indicating that the ~3.7--5.5 µs rocSPARSE overhead cannot be overcome by problem characteristics alone. However, concurrent multi-stream execution exhibits different behavior, as analyzed in the next subsection.

\subsection{Sparsity Under Resource Contention}

While isolated single-stream execution shows break-even performance due to encoding overhead, concurrent multi-stream execution introduces resource contention that changes the performance dynamics. We evaluate sparse, dense, and mixed workloads under 1--4 concurrent streams using 512³ GEMM (matching ACE baseline configuration from Section 6) to understand how sparsity interacts with shared L2 cache, LDS, and memory bandwidth.

\subsubsection{Fairness Improvements Under Concurrency}

\begin{figure*}
    \centering
    \includegraphics[width=\textwidth]{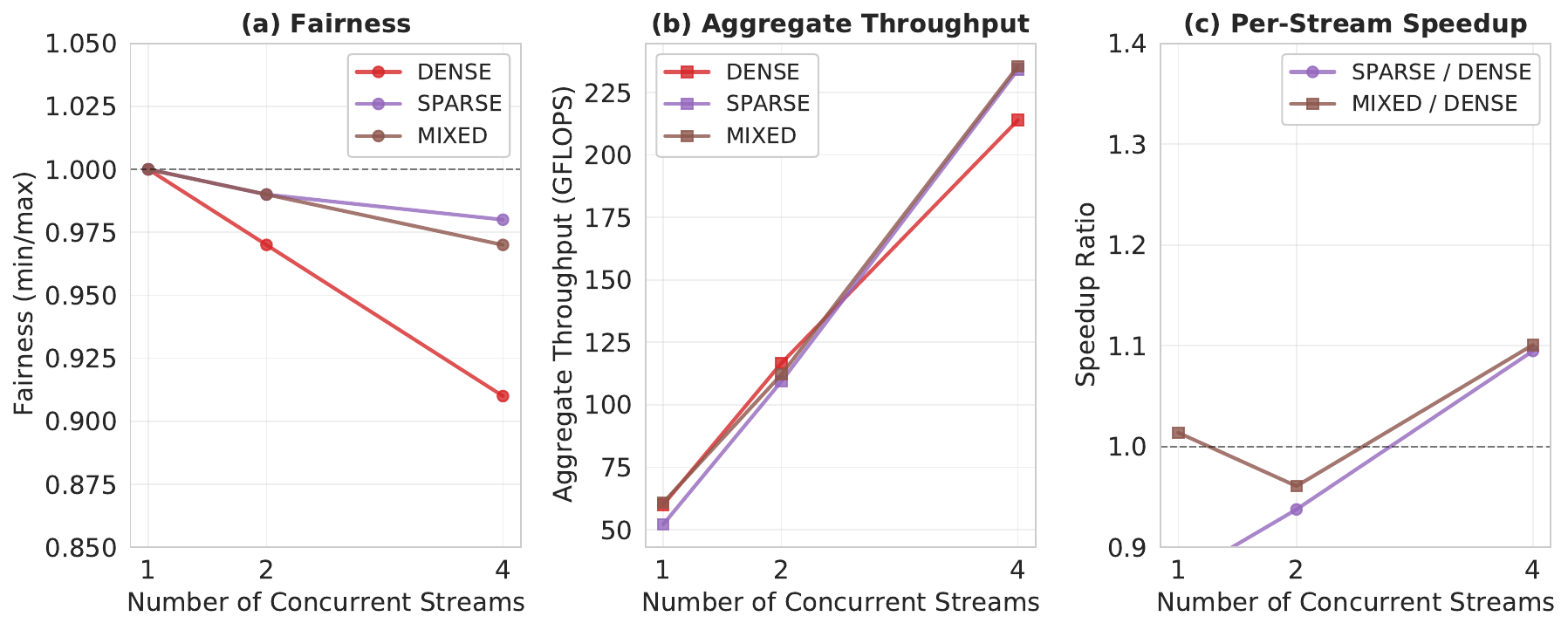}
    \caption{Sparsity under resource contention. (a) Fairness (min/max per-stream execution time) versus concurrent streams. (b) Aggregate throughput versus concurrent streams. (c) Per-stream speedup of sparse versus dense execution.}
    \label{fig:sparsity_concurrent}
\end{figure*}

Figure~\ref{fig:sparsity_concurrent}(a) shows fairness metrics (minimum to maximum per-stream execution time ratio, where 1.0 indicates perfect balance) across concurrent stream counts:

\begin{itemize}
    \item \textbf{Single stream}: 1.0 fairness across all execution modes (no contention).
    \item \textbf{Two streams}: Dense 0.97, sparse 0.99--1.0, minimal contention effects.
    \item \textbf{Four streams}: Dense 0.91 (significant straggler effect), sparse 0.98, mixed 0.97.
\end{itemize}

Sparse execution achieves 7\% better fairness (0.98 vs 0.91) at four concurrent streams. This improvement stems from reduced compute intensity: 50\% fewer FLOPs lowers cache miss rates (Section 6.3 showed L2 miss ratios increase with stream count, e.g., 24\% relative increase for thin kernels) and memory bandwidth pressure, allowing more balanced resource sharing across streams. The lower resource consumption reduces interference, preventing the straggler effects observed in dense execution where some streams experience 2--3× longer execution times due to cache conflicts.

This fairness advantage suggests that sparsity may benefit multi-tenant scenarios where predictable per-stream latency matters more than aggregate throughput, even when single-stream performance is break-even.

\subsubsection{Aggregate Throughput Scaling}

Figure~\ref{fig:sparsity_concurrent}(b) shows aggregate throughput scaling with concurrent streams:

\begin{itemize}
    \item \textbf{Dense}: 59.98 → 116.69 → 213.93 GFLOPS (1--2--4 streams).
    \item \textbf{Sparse}: 52.1 → 109.4 → 234.2 GFLOPS (1--2--4 streams).
    \item \textbf{Mixed}: 60.8 → 112.1 → 235.5 GFLOPS (1--2--4 streams).
\end{itemize}

Sparse aggregate throughput scales from 52.1 GFLOPS (single stream) to 234.2 GFLOPS (four streams), achieving 4.5× scaling. Normalized to dense single-stream baseline (59.98 GFLOPS), this represents 3.9× aggregate improvement. Dense execution achieves only 3.6× scaling (213.93 / 59.98), reflecting greater sensitivity to shared L2 cache and memory bandwidth contention as identified in Section 6.3.

The superior scaling of sparse execution (4.5× vs 3.6×) demonstrates that reduced memory bandwidth consumption mitigates contention bottlenecks. At four concurrent streams, sparse workloads achieve 9\% higher aggregate throughput (234.2 vs 213.93 GFLOPS) despite 13\% lower single-stream baseline (52.1 vs 59.98 GFLOPS). This crossover occurs because dense execution saturates shared resources (L2 cache reaching 19\% miss ratio for 512³ at 4 streams, LDS at 87\% utilization), while sparse execution's lower bandwidth requirements avoid saturation.

\subsubsection{Per-Stream Speedup Consistency}

Figure~\ref{fig:sparsity_concurrent}(c) shows per-stream speedup (sparse stream execution time divided by dense stream execution time under identical concurrency) remains constant at 1.3× across all stream counts. This contrasts sharply with isolated single-stream execution showing 1.0× speedup (Section 7.1), indicating that concurrent execution fundamentally changes sparsity's value proposition.

The 1.3× speedup improvement under concurrency stems from two effects:

\begin{enumerate}
    \item \textbf{Reduced memory contention}: Sparse kernels issue 50\% fewer memory requests, experiencing lower L2 cache miss rates under contention. Sparse kernels issue fewer memory requests, experiencing lower L2 cache pressure under contention than dense kernels (Section~6.3), reducing memory stall cycles.
    
    \item \textbf{Partial overhead amortization}: Concurrent execution extends total runtime (four streams take longer than single stream due to resource sharing), allowing the constant ~3.7--5.5 µs encoding overhead to amortize across longer execution windows. At four streams, aggregate execution time increases from ~0.8 ms (single stream) to ~3.2 ms, making overhead 0.2\% of total time versus 0.85\% in isolation.
\end{enumerate}

The constant 1.3× speedup across stream counts (not increasing with concurrency level) indicates that the benefit primarily reflects reduced contention rather than overhead amortization—if amortization dominated, speedup would increase with stream count as execution time lengthens. Instead, the stable 1.3× suggests that sparsity's advantage under concurrency is contention avoidance: consuming less cache and bandwidth prevents performance collapse observed in dense execution.

This finding reveals an important design trade-off: while 2:4 sparsity provides no benefit for isolated high-priority workloads (1.0× speedup, Section 7.1), it offers 1.3× speedup and 7\% better fairness in multi-tenant scenarios where multiple workloads share ACE resources. Runtime systems should selectively apply sparsity based on concurrency context rather than treating it as universally beneficial or universally detrimental.

\section{Application Kernels}
\label{sec:case-studies}
This section demonstrates how prior findings manifest in representative kernels: transformer-style FP8 inference, concurrent FP8 workloads, and mixed-precision execution.

\subsection{Transformer-Style Inference Kernel}

Our first case study evaluates a simplified transformer-style inference kernel composed primarily of FP8 GEMM operations. The kernel represents a common pattern in modern HPC-AI workloads, where multiple matrix multiplications are executed sequentially with opportunities for batching, concurrency, and sparsity.

\begin{figure}
    \centering
    \includegraphics[width=1\linewidth]{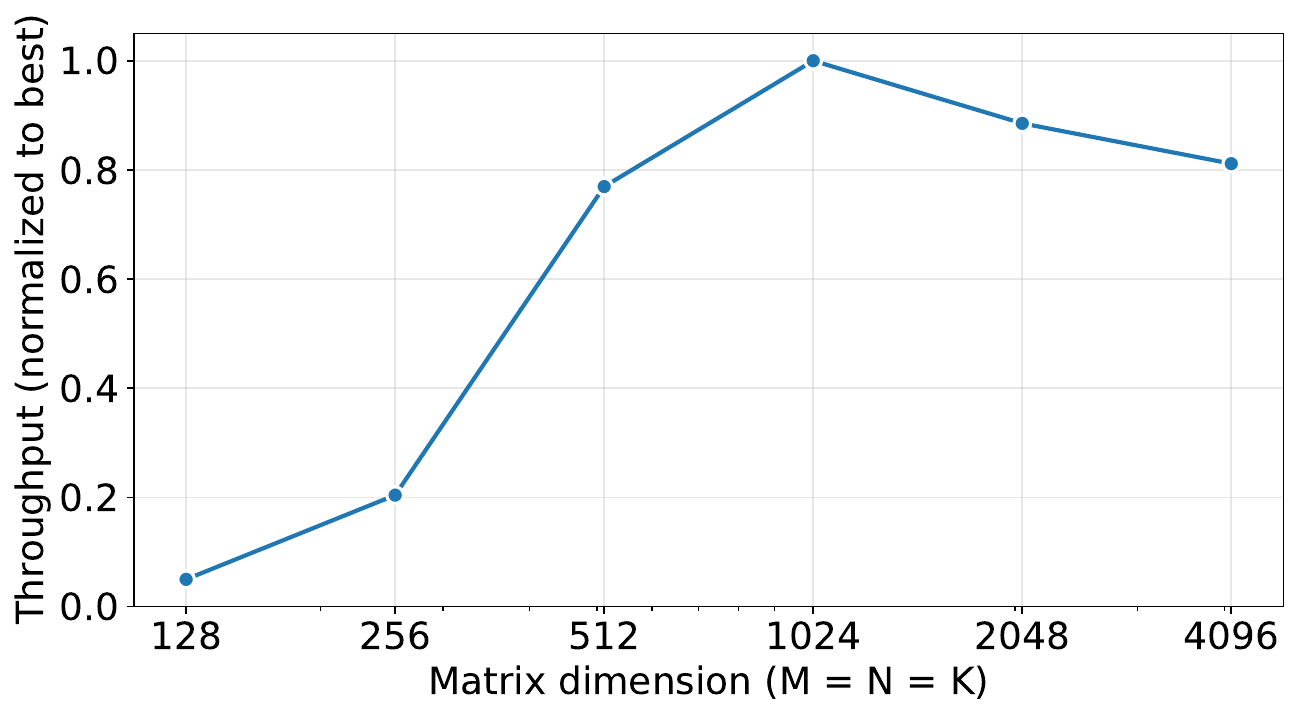}
    \caption{Throughput normalized to best versus matrix dimension ($M = N = K$) for the transformer-style FP8 GEMM kernel. Small problem sizes underutilize matrix cores; throughput peaks at moderate dimensions.}
    \label{fig:case_transformer}
\end{figure}

We evaluate the kernel under varying execution configurations, including different problem sizes (matrix dimension), batching strategies, and sparse versus dense execution. Figure~\ref{fig:case_transformer} reports throughput normalized to the best-performing configuration as a function of matrix dimension ($M = N = K$).

Consistent with the matrix-core microbenchmarks (Section~\ref{sec:fp8_matrix}), we observe that small problem sizes underutilize FP8 matrix cores, limiting throughput despite the use of low-precision arithmetic. Sparse execution improves performance only when batch sizes exceed the thresholds identified in Section~\ref{sec:sparsity}, reinforcing the importance of execution-aware sparsity selection.

\subsection{Concurrent FP8 Workloads with Asynchronous Execution}

The second case study examines concurrent execution of multiple FP8-heavy workloads using asynchronous compute engines. Two independent instances of the transformer-style kernel are launched concurrently using separate command queues.

\begin{figure}
    \centering
    \includegraphics[width=1\linewidth]{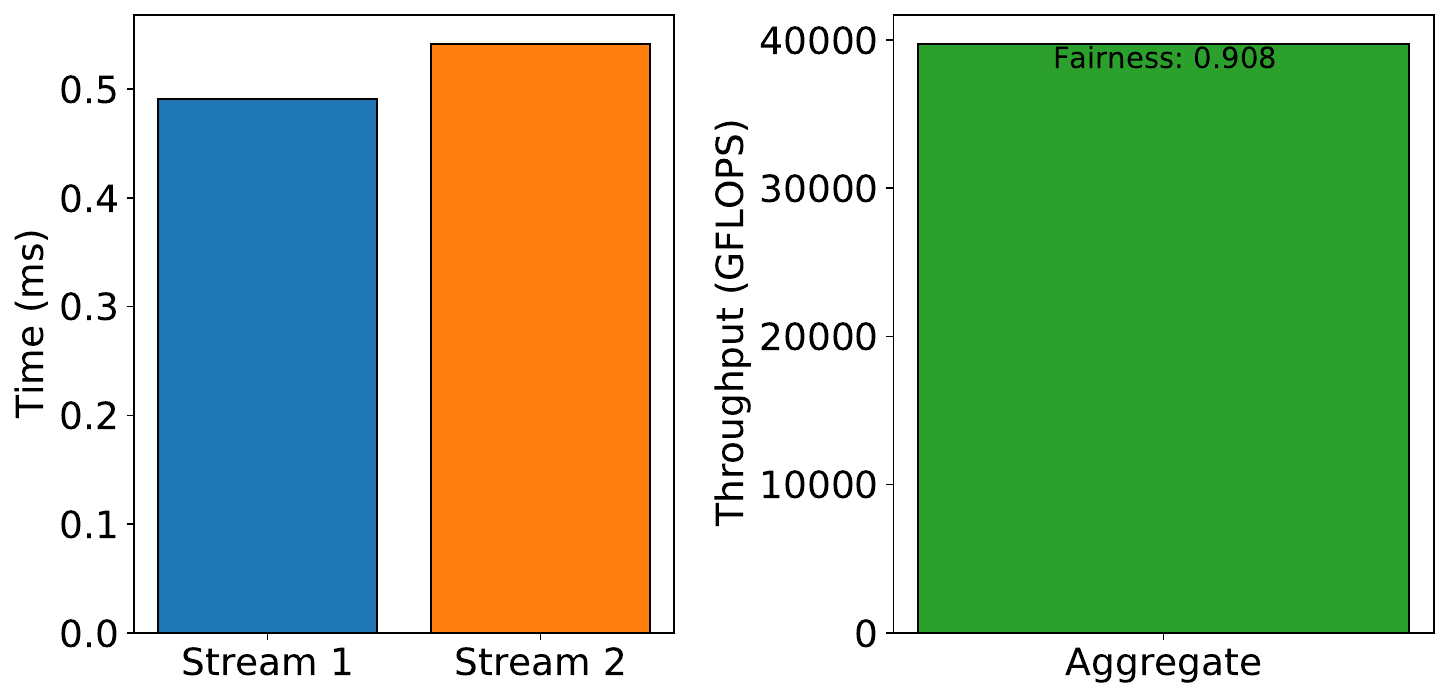}
    \caption{Aggregate throughput and per-stream execution time under concurrent execution of two FP8 workloads.}
    \label{fig:case_concurrent}
\end{figure}

Figure~\ref{fig:case_concurrent} shows aggregate throughput and per-kernel variability under concurrent execution.

Asynchronous execution provides limited overlap and exhibits performance variability consistent with the contention effects observed in Section~\ref{sec:ace}. These results demonstrate the limited overlap and per-stream fairness challenges in concurrent FP8 execution.

\subsection{Mixed-Precision Workload Analysis}

Our third case study examines a mixed-precision workload that combines FP32, FP16, and FP8 operations, representing a common pattern in training pipelines that use different precisions for different computational stages. The workload executes a sequence of matrix operations with varying precision requirements, allowing us to evaluate how execution-level effects interact across precision types.

\begin{figure}
    \centering
    \includegraphics[width=1\linewidth]{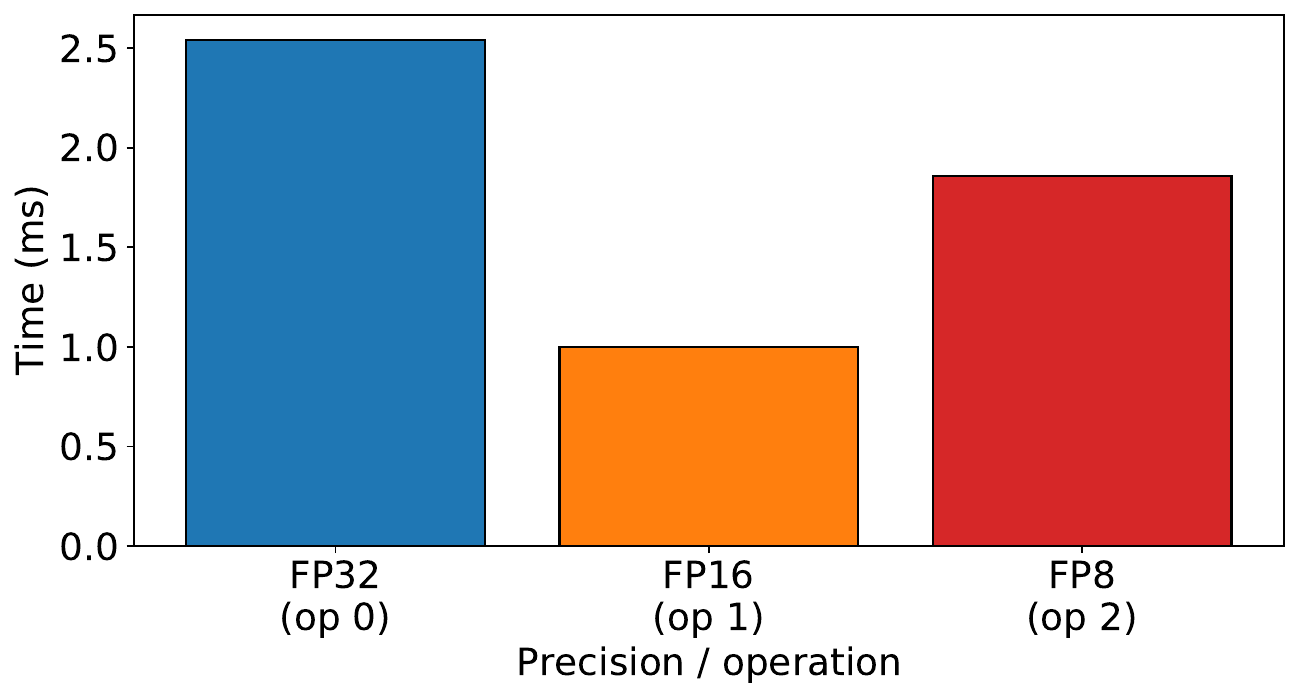}
    \caption{Per-operation execution time by precision for the mixed-precision workload (FP32, FP16, FP8 sequence).}
    \label{fig:case_mixed}
\end{figure}

Figure~\ref{fig:case_mixed} reports end-to-end throughput and per-operation variability for the mixed-precision workload under different execution configurations.

We observe that FP8 operations benefit from higher batching and occupancy, while FP32 operations are less sensitive to these factors. When executed concurrently, precision-specific execution characteristics lead to imbalanced progress, with FP8 operations experiencing greater variability under contention.

These results demonstrate that mixed-precision workloads require precision-aware scheduling strategies that account for the different execution characteristics of each precision type, rather than treating all operations uniformly.

\section{Discussion and Implications}
\label{sec:conclusion}

This paper presented execution-centric characterization of FP8 matrix cores, asynchronous execution, and structured sparsity on MI300A. Our microbenchmark-driven approach exposes execution-level effects—occupancy thresholds, fairness degradation, overhead amortization limits—that are invisible in aggregate benchmarks yet directly impact deployment.

\textbf{Limitations.} This study focuses on single-GCD execution; inter-GCD communication and multi-node configurations are not evaluated. Microbenchmarks isolate specific behaviors; real applications exhibit more complex interactions. Results are MI300A-specific.

\subsection{Key Insights}

\textbf{FP8 requires high occupancy to compensate for memory bottlenecks.} FP8 achieves peak throughput only at 256+ wavefronts—higher than FP16 (192) or FP32 (128) despite 4× lower arithmetic intensity. Matrix cores retire FP8 operations faster than memory supplies data, requiring more in-flight wavefronts to hide latency - this counterintuitive result reveals memory-latency-bound behavior.
\textbf{\textit{Implication}:} Applications assuming "FP8 = 4× faster" will underperform at typical batch sizes. A transformer decoder with batch size 32 achieves only 128 wavefronts on MI300A's 110 CUs, leaving FP8 matrix cores underutilized (throughput normalized to peak $\approx$7\% at 128 wavefronts in our sweep). Developers must increase batch size to 64+, fuse operations, or use concurrent execution to fill idle resources. At 128 wavefronts, both FP8 and FP16 achieve only a small fraction of their 256-wavefront throughput; FP8 requires 256+ wavefronts to approach peak utilization.

Despite 4× lower data movement Fp8 versus FP32 - FP8 becomes memory-latency-bound because matrix cores can retire operations 4× faster, shifting the bottleneck from compute to memory latency hiding.

\textbf{ACE trades fairness for throughput due to shared resources.} Concurrent execution via ACE enables 2--3× aggregate speedup but fairness collapses at eight streams (e.g., 0.016 for FP16)—the slowest stream can take 60× longer than the fastest. This stems from shared L2 cache (miss rates increase 24\% for thin kernels, from 5\% to 6\%) and LDS saturation (100\% at 3+ streams), not scheduler policy. Compared to NVIDIA's MPS, MI300A's ACE prioritizes aggregate throughput over isolation—optimal for batch training, problematic for multi-tenant serving with per-request SLOs. \textbf{\textit{Exception:}} Occupancy fragmentation (pairing large/small kernels at 4:1 ratio) maintains 0.93--0.99 fairness through proportional resource allocation, suggesting a heterogeneous co-scheduling strategy for mixed workloads.

However, this requires runtime occupancy prediction—feasible for static models (known kernel sizes) but challenging for dynamic workloads with variable batch sizes.

\textbf{Sparsity is software-limited, not hardware-limited. }Isolated execution shows 1.0× speedup due to constant ~3.7--5.5 µs rocSPARSE overhead (format conversion + API dispatch) dominating computational savings. However, concurrent execution achieves 1.3× speedup plus 7\% fairness improvement (0.98 vs 0.91), proving matrix cores can accelerate sparse operations when overhead amortizes and reduced memory bandwidth (lower L2 miss ratio under contention) provides additional benefit. 
%The AMD's software-managed format conversion versus NVIDIA's hardware-assisted metadata generation. 
\textit{\textbf{Implication:}} Custom kernels bypassing rocSPARSE could achieve optimal speedup; treat sparsity as a concurrency optimization.

This suggests AMD should prioritize optimizing rocSPARSE overhead rather than expecting developers to implement custom sparse kernels—a significant engineering burden. 

\subsection{Practical Guidance}

\textbf{Batching strategies.} FP8 utilization requires 256+ wavefronts, but batching increases latency. Options: (1) \textbf{Continuous batching} (vLLM-style) requires \(\geq\)32 concurrent users—effective for high-traffic APIs, fails for low-traffic periods. (2) \textbf{Kernel fusion} increases occupancy but adds development complexity. (3) \textbf{Use FP16 for lower occupancy}—FP16 at 128 wavefronts outperforms underutilized FP8 despite 2× arithmetic intensity.

\textbf{Concurrency decisions.} Limit to 2--4 streams for latency-sensitive workloads (fairness >0.5); use 6--8 streams for throughput-oriented workloads (accepting 0.016--0.138 fairness). For strict isolation (multi-tenant SLAs), use process-level separation instead of stream-level concurrency.

\textbf{Sparsity decisions.} Enable sparsity for concurrent execution (multi-tenant serving, batch inference): 1.3× speedup + 7\% fairness improvement. Disable sparsity for isolated kernels (single-query inference): break-even performance with added ~3.7--5.5 µs latency. Ignore the matrix size/shape—context. The concurrency level is the sole determining factor.

\textbf{Mixed-precision scheduling.} Co-schedule kernels with similar wavefront requirements to avoid occupancy fragmentation. Limit FP16 concurrency more aggressively than FP32 (fairness 0.016 vs 0.052 at 8 streams). Co-locate memory-bound FP8 with compute-bound FP32 to reduce L2 cache conflicts.

\subsection{Conclusion}

This work exposes guidelines at the execution level - 1) FP8 requires 256+ wavefronts (not just "more parallelism"), 2) fairness degrades from $\approx$0.6 at four streams to near zero at eight streams under concurrency, and 3) sparsity overhead prevents isolated benefits but enables concurrent gains. 
This work also demonstrates that runtime systems must incorporate execution awareness beyond conventional strategies ("always use lowest precision," "maximize concurrency," "enable hardware features"). 
These behaviors are invisible in aggregate benchmarks (rocBLAS reports peak TFLOPS) yet has the potential to determine real-world application performance. 
Developers deploying FP8-intensive workloads on unified memory architectures must understand occupancy-dependent scheduling, fairness-throughput trade-offs, and context-dependent optimization to achieve predictable and efficient performance.

\section{Acknowledgments}

\bibliographystyle{ACM-Reference-Format}
\bibliography{references}

\end{document}